\documentclass[twocolumn,twoside,slac_two]{revtex4}
\usepackage{graphicx}
\usepackage{fancyhdr}
\newcommand{\be}{\begin{equation}}
\newcommand{\ee}{\end{equation}}
\newcommand{\bea}{\begin{eqnarray}}
\newcommand{\eea}{\end{eqnarray}}
\newcommand{\bean}{\begin{eqnarray*}}
\newcommand{\eean}{\end{eqnarray*}}
\newcommand{\qq}{$q\bar{q}$}
\newcommand{\cc}{$c\bar{c}$}
\newcommand{\cs}{$c\bar{s}$}
\newcommand{\cu}{$c\bar{u}$}
\newcommand{\cd}{$c\bar{d}$}

\begin{document}

\title{Rumsfeld Hadrons}


\author{F.E.Close}
\affiliation{Rudolf Peierls Centre for Theoretical Physics;
University of Oxford; Oxford OX1 3NP; England}

\begin{abstract}

Donald Rumsfeld, in attempting to excuse the inexcusable, once (in)famously said that 
``there are things that we know we know; there are things that we know we don't know; 
and then there are things that we don't know that we don't know". Recent discoveries
about hadrons with heavy flavours fall into those categories. It is of course the third
category that is the most tantalising, but lessons from the first two may help resolve the third.

\end{abstract}

\maketitle

\thispagestyle{fancy}


\section{Things that we know we know}

We have heard reported observation of the $B_c$ with mass $m(B_c) = 6276.5 \pm 4.0 \pm 2.7$MeV 
and comparison of its mass with
predictions from various models and lattice QCD\cite{bctalk}.
Not everything is mysterious. Compare this lightest $b\bar{c}$ with $(m(\psi)+ m(\Upsilon))/2 = %%@
6278.6$. 
They agree to better than a part per mille. This illustrates how apparent agreements with the %%@
mass are driven by the large
intrinsic mass scales of the $b$ and $c$ and that yet again the mass scales of hadrons are 
phenomenologically rather straightforward. The interesting dynamics will come when excitations of %%@
the $b\bar{c}$ are found.

A more profound testing of QCD effects has come from the discovery of $\Sigma_b$ and $\Sigma_b^*$ %%@
at CDF\cite{cdftalk}.
The chromomagnetic splittings between baryons were predicted thirty years ago\cite{dgg}. The %%@
$\Delta - N$ splitting of 300MeV
involves gluon exchange among all three constituent flavours. If one of these were replaced by an %%@
infinitely massive flavour,
its contribution to the colour magnetism would vanish as magnetic couplings are inversely %%@
proportional to mass. The spin couplings
of the two light flavours then leave a residual splitting of 200MeV between the light $\Lambda_Q$ %%@
and the (degenerate) $\Sigma_Q$
and $\Sigma_Q^*$.

In reality we dont  have infinitely massive flavours but can compare the trend as $Q$ is in turn %%@
$s,c,b$
\cite{fecbook}. The $\Sigma - \Lambda$
separation grows while the $\Sigma-\Sigma^*$ come together. The trend is well known for strange %%@
and charm; the CDF results for
bottom confirm this beautifully. Although we do not fully understand what the degrees of freedom %%@
that we call
``constituent quarks" are, nonetheless, they behave in a remarkably simple fashion. It remains a %%@
challenge to theory to explain why.
(See also \cite{kl07} for extensive discussion about this area).

The area that is phenomenologically understood extends to mesons where the \qq~ are in relative
$L=0$. In the recent past there has been much new information about the $c\bar{s}$ states. 
The predictions and observations
are consistent for the $0^-;1^-$ states (for an up to date comparison see \cite{cs05,DsGeneral}). 

BaBar recently announced the discovery of a new $D_s$ state seen in $e^+e^-$ collisions 
decaying to $D^0K^+$ or $D^+K^0_S$\cite{palano}. The 
Breit-Wigner mass of the new state is 

\be
M(D_{sJ}(2860)) = 2856.6 \pm 1.5 \pm 5.0 \ {\rm Mev}
\ee
and the width is
\be
\Gamma(D_{sJ}(2860)) =  48 \pm 7 \pm 10\ {\rm MeV}.
\ee
 There is no evidence of the $D_{sJ}(2860)$ in the $D^*K$ decay mode\cite{palano} 
or the $D_s \eta$ mode.
There is, furthermore, structure in the $DK$ channel near 2700 MeV that 
yields Breit-Wigner parameters of

\be
M(D_{sJ}(2690)) = 2688 \pm 4 \pm 2 \ {\rm MeV}
\ee
and
\be
\Gamma(D_{sJ}(2690)) = 112 \pm 7 \pm 36 \ {\rm MeV}.
\ee

The state $D_s(2690)$ 
has the characteristics of a vector and is consistent with
being the $2S(^3S_1)$ or possibly mixed with $^3D_1$. Ref\cite{cs05} discusses this in more %%@
detail. If this is true, then
a test is to produce the state in $B$ decays.

The production of the radially excited $D_{s0}$ in $B$ decays can be estimated with the ISGW %%@
formalism\cite{chris}. 
Since vector and scalar $c\bar s$ states can be produced directly from the $W$ current, the %%@
decays
$B \to D_s^*(2S) D_{(J)}$ or $D_{s0}(2P) D_{(J)}$ serve as a viable source excited $D_s$ states. 
Computationally, the only
differences from ground state $D_s$ production are kinematics and the excited $D_s$ form factors.
The relative rates for excited vector production were given in Ref. \cite{cs05,cs06}:

\begin{equation}
B \to D_s^* \bar D : B \to D_s^*(2S) \bar D \approx 1 : 0.3 - 0.7.
\end{equation}

In the $L=1$ sector the model predicts the radial excitation $2P(^3P_0)$ at 2820. The observation %%@
of $D_{sJ}(2860)$
is consistent with this though higher spin $3^-$ has also been suggested\cite{3-}. Angular %%@
distributions can sort this out.
If it is $0^+$ then some interesting further tests can ensue which touch on the next level of %%@
conundrum:

\section{Things that we know we dont know}

In the $L=1$ \cs~ sector, why are the masses of $0^+(2317)$ and $1^+(2460)$ so much lower than %%@
the models had expected?
I have always felt that this is an example of where the naive quark model is too naive; for %%@
details see my summary
talk in Hadron03\cite{hadron03}. When a \qq~ state occurs in $L=1$, but can couple to hadron %%@
pairs in S-wave, the latter
will distort the underlying \qq~ picture. The \cs~ $0^+$ state predicted above $DK$ threshold %%@
couples to the $DK$ continuum.
This mixes $DK$ into the wavefunction, and leads to a weakly bound quasi-molecular state just %%@
below $DK$ threshold.
Analogous dynamics was predicted also for the $1^+$ coupling to $D^*K$. 
This effectively ``unquenches" the quark model for such states
and led to the molecular picture of \cite{BCL}. This has been discussed extensively and the %%@
effects of the intrinsic \cs~
seed also investigated in \cite{beveren}.

There has also been suggestion that this is an effect of chiral symmetry\cite{chiral}, the $0^- - %%@
1^-$ states with 143 MeV mass 
gap matching their $0^+ - 1^+$ chiral partners with 146MeV. This is intriguing but among things %%@
that I don't undertstand are: 
(i) why does this apply for \cs~ where no manifestly light flavours are involved rather than \cu~ %%@
or \cd~ ; 
(ii) which of the two $1^+$ states is the comparison supposed to be made with? (In the infinitely %%@
massive quark 
limit there is a clear answer, but it is trivial as the splittings go to zero; for finite masses %%@
there is mixing 
between the $S=0,1$ basis and the $p_{1/2} - p_{3/2}$ states and the comparison is non-trivial).

The full spin-dependent structure expected at order
$\alpha_s^2$ in QCD has been computed by Pantaleone {\it et al.}\cite{tye} and reveals that
an additional spin-orbit contribution to the spin-dependent interaction exists when quark masses 
are not equal. When these are incorporated 
in a constituent quark model there can be significant mass shifts leading to a lowered mass for %%@
the
$D_{s0}$ consistent with the $D_{s0}(2317)$\cite{LS}.
The identification of analogue states with bottom flavour could help decide among these pictures.

If both the 2860 and 2317 are canonical \cs~ $0^+$ states in $2P$ and $1P$ respectively, then %%@
their relative
production in $B$-decays should be\cite{cs06}:

\begin{equation}
B \to D_s(2860) \bar D : B \to D_s(2317) \bar D \approx 1 : 0.6 - 1.8.
\end{equation}
A comparison in $B$-decays is warranted.

\section{Things that we don't know we don't know}

There is much unexpected activity showing up in the charmonium sector: $X(3940)$ seen in
$\psi\omega$~\cite{bellepsiomega}, $Y(3940)$ seen in
$D^\ast\bar{D}$~\cite{belleddstar}, $\chi_{c2}(3930)$~\cite{bellechi2}
and $Y(4260)$~\cite{babar}. (The inclusion of charge-conjugated
reactions is implied throughout.) Furthermore there are also
three prominent enhancements
$X$ in $e^+e^- \to \psi + X$~\cite{belleddstar}, which are
consistent with being the $\eta_c,\eta_c'$ and $\chi_0$.

In $e^+e^- \to \psi + X$ there is no sign of the $X(3872)$; this state now appears to have
$C=+$ and be consistent with $1^{++}$~\cite{3872data}. This $J^{PC}$
was first suggested in Ref.~\cite{fcpage3872} and a dynamical picture of it
as a quasi-molecular $D^{\ast 0}\bar{D}^0$ state discussed
in Refs.~\cite{fcpage3872,swanson3872}. The suppression of this state among
prominent $C=+$ charmonium states~\cite{belleddstar} may thus be
consistent with its molecular versus simple \cc~ nature.

 Why the $\chi_0$ is prominent in $e^+e^- \to \psi + X$ and the $\chi_2$ not is one question 
(contrast with light flavours where  $e^+e^- \to \omega f_2$ is
clearly seen, in particular in $\psi$ decay). The nature of the structure $X(3940)$ is clearly %%@
another
question. On mass grounds it could contain radially excited $\chi$ states; 
but if so why does it not appear in $D\bar{D}$? (If $\chi_0$ is
prominent in the data on $e^+e^- \to \psi + X$, 
then if $2P$ states are strongly produced, one would expect 2P($\chi_0$) also to be significant.) 
Given that $\eta_c(2S)$ is prominent, then perhaps $\eta_c(3S)$ is also, which
would explain the suppressed $D\bar{D}$. A third possibility is that hybrid charmonium with C=+
is being produced, which also is predicted not to decay to $D\bar{D}$.

There is a folklore that $X(3940)$ is too light to be hybrid charmonium, but I disagree with
that. Lattice QCD and flux tube models agree that the typical mass scale for an exotic $1^{-+}$
is $\sim $4.2 GeV\cite{ukqcd,bcs}; 
given that spin-dependent splittings place a $0^{-+}$ lighter than this and $1^{--}$ slightly %%@
heavier\cite{bcd}, it is
quite plausible that a $0^{-+}$ hybrid (mixing with $\eta_c(3S)$ ?!) is in this region, and that %%@
the vector $Y(4260)$ is 
also part of this story. Proving this will be hard though; I shall return to this later.

First let's consider the lightest of the novel charmonium states, the $X(3872)$ at $D^0D^{*0}$
threshold and which is almost certainly an axial meson. This has been known for some time and is %%@
generally agreed to have a 
tetraquark affinity; whether it is a genuine $D^0D^{*0}$ molecule or
a compact $cu\bar{cu}$ is a more subtle issue. If the quark-pairs are tightly clustered into %%@
di-quarks, then a $S=0$
and $S=1$ are required to make the $1^{++}$. Consequently other states, combinations of $0^+-0^+$ %%@
and $1^+-1^+$
would be expected. The absence of such a rich spectrum suggests that the overriding dynamics is %%@
that the constituents
rearrange into loosely bound colour singlet $c\bar{u}$-$u\bar{c}$, or $D^0D^{*0}$.
I don't plan to discuss that today but would raise
a couple of points about the binding mechanism that seem not to have been widely considered.

There is a significant possibility that this state and the ``conventional" $\chi_1(3500)$ may %%@
have
some mixing. If so there will be some isospin breaking in the latter's decays such that 
$br(\chi_1 \to K^+K^-\pi) > br(K^0\bar{K^0}\pi)$.  

It is the dynamical origin of this state that concerns me more. The arguments for it being a %%@
molecule were discussed in 
ref\cite{cp04} and $\pi$ exchange as a binding mechanism was also considered. 
Tornqvist had earlier suggested that a whole series of hadrons 
might form such ``deuson" states via $\pi$ exchange\cite{torn}. 
Swanson\cite{swanson3872} considered a quark exchange model, in which the 
coincidence of $DD^*$
and $\psi\omega$ as well as $\psi \rho$ energies played a role. Swanson found an attractive force %%@
but also showed that, 
within the approximations employed, this was insufficient to bind and he had
to include $\pi$ exchange to do so; as such this effectively recovers Tornqvist's $\pi$-exchange %%@
model
and leaves quark-interchange having little to do with the binding. There is
one comment however: Swanson considered only short distance (contact) gluon exchange in his quark
exchange calculation and not the contribution of the tensor force that gluons also induce. Note %%@
that
the Yukawa $\pi$ exchange potential also is not sufficient to bind; it is the tensor force that %%@
is found to be 
essential\cite{torn,swanson3872}. So it would be interesting to see what happens if the 
tensor contribution from gluon exchange is also included; this is being investigated 
by C Thomas\cite{cthom}. Understanding this may also have implications for other anomalous %%@
charmonium
states. It would also have potential implications for the presence or absence of analogous
states involving heavy flavours, and also for $D_sD_s^*$: this would receive no contribution from
$\pi$ exchange, and $\eta$ exchange is expected to be negligible whereas quark-interchange could %%@
occur. Furthermore,
$\pi$ exchange can also occur between $D$ and $D^*$ (i.e with no $\bar{D}$ \cite{torn}) whereas %%@
quark exchange
would not link to the $\psi\omega$ and the effects would generally differ.

I now turn to hybrid charmonium and evaluate the prospects that it is being exposed. There are
three states of interest (i) $X(3940)$ in the recoil spectrum  $e^+e^- \to \psi + X(3940)$, which
is not seen in $\omega \psi$; (ii) $Y(3940)$ seen in B-decay and which is seen in $\omega \psi$;
(iii) $Y(4260)$ which is $1^{--}$ in $e^+e^- \to \psi \pi\pi$, with no observed decay into %%@
$D\bar{D}$.
The fact that there is no sign of established
3S/2D(4040/4160)
4S(4400)
in the $\psi\pi\pi$ data already marks this state as anomalous and eliminates
conventional explanations as potential states are already apparently occupied.
The mass, large width into $\psi\pi\pi$, small leptonic width ($O(5-80)$eV, contrast
$O$(keV) for known states), affinity for $DD_1$ threshold and
apparent decay into $\psi \sigma$ or $\psi f_0(980)$ are all consistent with predictions made for
hybrid vector charmonium\cite{cp05}. I now assess the empirical status of hybrid charmonium and
other possible interpretations of the states.

 The eight low-lying hybrid charmonium states ($c\bar{c}g$) were
predicted in the flux-tube model to occur at $4.1-4.2$ GeV~\cite{bcs},
and in UKQCD's quenched lattice QCD calculation with infinitely heavy
quarks to be $4.04\pm 0.03$ GeV (with un-quenching estimated to raise
the mass by $0.15$ GeV)~\cite{ukqcd}.  The splittings of $c\bar{c}g$
from the above spin-average were predicted model-dependently for long
distance (Thomas precession) interactions in the flux-tube
model~\cite{merlin}, and for short distance
(vector-one-gluon-exchange) interactions in cavity
QCD~\cite{bcd,pthesis}. For the $1^{--}$ state the long and short
distance splittings respectively are $0$ MeV and $60$ MeV.
 Long ago the spin-dependent mass shifts were calculated in
cavity QCD, though the resulting pattern is expected to be more generally true\cite{bcd,drum}. 
Quenched lattice QCD indicates that the $c\bar{c}g$
$1^{--},\; (0,1,2)^{-+}$ are less massive than
$1^{++},\; (0,1,2)^{+-}$~\cite{juge}. The spin splitting for this lower
set of hybrids in quenched lattice NRQCD is
$0^{-+}<1^{-+}<1^{--}<2^{-+}$~\cite{drum}, at least for $b\bar{b}g$.
This agrees with the ordering found in the model-dependent
calculations for $q\bar{q}g$~\cite{bcd} in the specific case of
$c\bar{c}g$~\cite{pthesis,merlin}. For $b\bar{b}g$ lattice QCD
predict substantial splittings $\sim 100$ MeV or greater~\cite{drum},
which become even larger in the model-dependent calculations for
$c\bar{c}g$~\cite{pthesis,merlin}.  

Thus the consensus is that the resulting pattern
is, in decreasing mass, $1^{--}; 1^{-+}; 0^{-+}$ with the mass gap between each state being 
the same and of the order
of 10-100MeV. Thus theory strongly
indicates that if $Y(4260)$ is $c\bar{c}g$, and the splittings are not
due to mixing or coupled channel effects, then the $J^{PC}$ exotic
$1^{-+}$ and non-exotic $0^{-+}$ $c\bar{c}g$ are below
$D^{\ast\ast}\bar{D}$ threshold, making them narrow by virtue of the
selection rules. The $1^{-+}$ decay modes~\cite{dunietz} and branching
ratios~\cite{cgodfrey} have extensively been discussed. Thus it is consistent to identify %%@
possible states as 
$1^{--}(4.25); 1^{-+}(4.1);0^{-+}(3.9)$ and to speculate whether there are two states
$1^{-+}(4.1);0^{-+}(3.9)$ in either the $X/Y(3940)$ structures of Belle or $e^+e^-$. 
This is clearly a question that statistics from a super-B factory may resolve for the
B-decays or $e^+e^- \to \psi + X$.

Mass arguments alone will not be convincing; we need to understand the dynamics of production and
decay and show that these fit best with hybrid states.

A lattice inspired flux-tube model showed that the decays of hybrid
mesons, at least with exotic $J^{PC}$, are suppressed to pairs of
ground state conventional mesons~\cite{ipaton,ikp}.  This was extended
to all $J^{PC}$, for light or heavy flavours in Ref.~\cite{cp95}.
A similar selection rule was found in constituent gluon
models~\cite{pene}, and
their common quark model origin is now understood~\cite{pagesel}.
It was further shown that these selection rules for light flavoured
hybrids are only approximate, but that they become very strong for
$c\bar{c}$~\cite{cp95,pthesis}.  This implied that decays into
$D\bar{D},\: D_s\bar{D}_s,\: D^\ast \bar{D}^\ast$ and $D_s^\ast
\bar{D}_s^\ast$ are essentially zero while $D^\ast\bar{D}$ and
$D_s^\ast\bar{D}_s$ are very small, and that $D^{\ast\ast}\bar{D}$, if above
threshold, would dominate. (P-wave charmonia are
denoted by $D^{\ast\ast}$).  As $c\bar{c}g$ is predicted around the
vicinity of $D^{\ast\ast}\bar{D}$ threshold, the opportunity for
anomalous branching ratios in these different classes was proposed as
a sharp signature~\cite{cp95,bcs}. (To the best of our knowledge
Ref.~\cite{cp95} was the first paper to propose such a distinctive
signature for hybrid charmonium.)

It has become increasingly clear recently that there is an affinity
for states that couple in S-wave to hadrons, to be attracted to the
threshold for such channels~\cite{hadron03}. The hybrid candidate
$1^{--}$ appearing at the S-wave $D_1(2420)\bar{D}$ is thus interesting.

 More recently the signatures for hybrid charmonia were expanded to
note the critical region around $D^{\ast\ast}\bar{D}$ threshold as a
divide between narrow states with sizable branching ratio into
$c\bar{c}\; +$ light hadrons and those above where the anomalous
branching ratios would be the characteristic
feature~\cite{dunietz,cgodfrey}.  Here
widths of order 10 MeV were anticipated around the threshold.  It
was suggested to look in $e^+e^-$ annihilation in the region
immediately above charm threshold for state(s) showing such anomalous
branching ratios~\cite{cgodfrey}. The leptonic couplings
 to $e^+e^-,\; \mu^+\mu^-$ and $\tau^+\tau^-$  were expected
to be suppressed~\cite{ono}
(smaller than radial S-wave $c\bar{c}$ but larger
than D-wave $c\bar{c}$, but with some inhibition due to the fact that
in hybrid vector mesons spins are coupled to the $S=0$, whose coupling
to the photon is disfavoured~\cite{cgodfrey}). 

 The dominant mode
would be to $DD_1$ or $D^*D_0$ if kinematically allowed; these being S-wave and near threshold,
with low recoil momentum, there can be a significant amplitude for their constituents to %%@
rearrange leading to the
kinematically allowed decay channels $\psi \pi\pi$ including $\psi f_0/\sigma$. This would then %%@
be the explanation
of the strong decay width to $\psi \pi\pi$, which would otherwise superficially be OZI 
suppressed.

There are several of the theoretical expectations already given for
$c\bar{c}g$ that are born out by $Y(4260)$: (1) Its mass is
tantalizingly close to the prediction for the lightest hybrid
charmonia; (2) The expectation that the $e^+e^-$ width should be
smaller than for S-wave $c\bar{c}$ is consistent with
the data\cite{cp05};
%(3) The idea that $c\bar{c}\; +$ light hadrons have a substantial
%width agrees with Eq.~\ref{pipi}.
(3) The predicted affinity of hybrids to $D^{\ast\ast}\bar{D}$
could be related to the appearance of the state near the
$D^{\ast\ast}\bar{D}$ threshold. The formation of $D^{\ast\ast}\bar{D}$
at rest may lead to significant re-scattering into $\psi\pi^+\pi^-$,
which would feed the large signal.

 The nearness of $Y(4260)$ to the $D_1(2420) \bar{D}$ threshold, and
to the $D_1' \bar{D}$ threshold, with the broad $D_1'$ found at a mass
of $\sim 2427$ MeV and width $\sim 384$ MeV~\cite{dmass}, indicate
that these states are formed at rest. Also, these are the lowest open
charm thresholds that can couple to $1^{--}$ in S-wave (together with
$D_0 \bar{D}^\ast$, where the $D_0$ mass $\sim 2308$ MeV and
width $\sim 276$ MeV~\cite{dmass}).  Flux-tube model predictions are
that the D-wave couplings of $1^{--}\; c\bar{c} g$ to the $1^{+}$ and
$2^{+}$ $D^{\ast\ast}$ are small~\cite{cp95,pthesis,pss}; and there is
disagreement between various versions of the model on whether the
S-wave couplings to the two $1^{+}$ states are large. If these
couplings are in fact substantial, the nearness of $Y(4260)$ to the
thresholds may not be coincidental, because coupled channel effects
could shift the mass of the states nearer to a threshold that it
strongly couples to; and it would experience a corresponding
enhancement in its wave function. The broadness of $Y(4260)$ also
implies that its decay to $D_1(2420) \bar{D},\; D_1' \bar{D}$ and
$D_0(2308)\bar{D}^\ast$ which feed down to $D^\ast \bar{D} \pi$ and $D
\bar{D} \pi$~\cite{cs05} would be allowed by phase space and
should be searched for to ascertain a significant coupling to
$D^{\ast\ast}$.

 Flux-tube model width predictions for other charm modes
are $1-8$ MeV for $D^\ast \bar{D}$~\cite{pss},
with $D\bar{D},\: D_s\bar{D}_s,\:
D^\ast \bar{D}^\ast$ and $D_s^\ast \bar{D}_s^\ast$ even more suppressed.
 Thus a small $D\bar{D}$ and $D_s\bar{D}_s$ mode could single out the hybrid
interpretation.
Unless there is significant re-scattering from $\psi f_0(980)$,
the hybrid decay pattern is very different from the $c\bar{s}s\bar{c}$
four-quark interpretation for $Y(4260)$ which decays predominantly in
$D_s \bar{D}_s$~\cite{maiani}. Thus data on the latter channel,
or limit on its coupling, could be a
significant discriminator for the nature of this $Y(4260)$.

The data\cite{eecharm} on $e^+e^- \to D_s \bar{D_s}$ show
a peaking above threshold around 4 GeV but no evidence of affinity for a structure at 4.26GeV.
This is suggestive but needs better quantification. 
If these data are confirmed, then as well as ruling out a $cs\bar{cs}$ at this mass, they will %%@
also
add support to the hybrid interpretation. 
The same data also show there is no significant coupling of $Y(4260)$ to $D\bar{D}; D^*\bar{D}$ %%@
or $D^*\bar{D^*}$, 
all of which are in accord with predictions for a hybrid state.

Before finally concluding that the $Y(4260)$ is hybrid charmonium, we must eliminate a third %%@
possibility:
is the $1^{--}$ $X(4260)$
 an effect of $\pi$ exchange attraction near the $DD_1$ threshold? Note that this is the
first threshold in $e^+e^-$ annihilation to charm where the charmed mesons emerge in
S-wave. This would be
analogous to the $1^{++}$ $X(3872)$ $DD^*$ that we discussed earlier. In the $Y(4260)$ case,
$\pi$-exchange connects $DD_1 \to D^*D_0$ and gives attraction in the $1^{--}$, $I=0$
channel (the $Y(4260?)$) and in the $1^{-+}$ $I=1$ channel (a truly exotic beast!).
There is also the question of whether quark-exchange gives attraction. If so this would
potentially allow enhancement near the $D_s D_{s1}$ threshold, whereas neither $\pi$-exchange
nor s-channel resonances would expect such a structure.

An indirect hint that $Y(4260)$ might be connected to resonant gluonic excitation is that
there appears to be an analogous phenomenon in the $s\bar{s}$ sector\cite{ss,2175}. 
The cross section for $e^+e^- \to K^+K^-\pi^+\pi^-$ has significant contribution from $e^+e^- \to %%@
KK_1$
with rescattering into $\phi\pi\pi$. A resonance with width $\Gamma = 58 \pm 16 \pm 20$MeV with
large branching ratio into $\phi \pi\pi$ is seen with mass of 2175MeV\cite{2175}. 
Simple arithmetic shows that the mass gap from this state to 
$m(\phi)$ is within the errors identical to that between $Y(4260)$ and $m(\psi)$. This is perhaps
reasonable if the cost of exciting the gluonic flux-tube is not sensitive to the masses of the
\qq~ involved (as lattice QCD seems to suggest), in which case a hybrid vector production and %%@
decay
is consistent with data. The $KK_1$ and $K^*K_0$ thresholds do not relate so readily to the 2175 %%@
state
as do the analogous charm states with the 4260, which makes it less likely perhaps that the 4260 %%@
and 2175 can be
simply dismissed as non-resonant effects associated with S-wave channels opening. While it may %%@
not
be possible to prove that these are
definitively signals for hybrid vector mesons, all of the phenomena are in accord with %%@
expectations for hybrids.

If this 4260 state is not hybrid vector charmonium, then where it is?

Suppose that it is. Where else should we look? Clearly the $[0,1]^{-+}$ states predicted to lie %%@
below
the $Y(4260)$ become interesting. The properties and search pattern for such states are discussed %%@
in ref.\cite{cgodfrey}.
In $e^+e^- \to \psi + X$ it is possible that such states could feed the signal at 3940MeV. If the
production is via strong flux-tube breaking there is a selection rule\cite{cburns} that %%@
suppresses
$\psi + X$ when $X$ has negative parity. However, it is possible that the dominant production for %%@
\cc~ + \cc~
is by ``preformation", where a perturbative gluon creates the second \cc~ pair (the highly %%@
virtual photon
having created the initial pair). In such a case there is no selection rule forbidding $X \equiv %%@
[0,1]^{-+}$ hybrids;
however, the amplitude will be proportional to the short distance wavefucntion of the hybrid, %%@
which is expected to be small
compared to those of e.g. $\eta_c(3S)$ though perhaps comparable to those of $\chi_J$. 
Thus it would be interesting to measure the
$J^{PC}$ of the $X(3940)$ region to see if it contains exotic $1^{-+}$. If such a signal 
were found then we could truly be sure that
hybrid charmonium had been revealed.

\begin{acknowledgments}

This work is supported,
in part, by grants from
the Particle Physics and
Astronomy Research Council, and the
EU-TMR program ``Eurodice'', HPRN-CT-2002-00311.

\end{acknowledgments}

\bigskip % extra skip inserted
%\begin{thebibliography}{9}   % Use for  1-9  references

\end{document}